\documentclass{vgtc}                          
\usepackage[bookmarks=false]{hyperref}

\ifpdf
  \pdfoutput=1\relax                   
  \usepackage{graphicx}                
  \DeclareGraphicsExtensions{.pdf,.png,.jpg,.jpeg} 
\else
  \usepackage{graphicx}                
  \DeclareGraphicsExtensions{.eps}     
\fi%

\graphicspath{{figures/}{pictures/}{images/}{./}} 

\usepackage{microtype}                 
\PassOptionsToPackage{warn}{textcomp}  
\usepackage{textcomp}                  
\usepackage{mathptmx}                  
\usepackage{times}                     
\usepackage{cite}                      
\usepackage{tabu}                      
\usepackage{booktabs}                  

\onlineid{1288}

\vgtccategory{Methodological}

\vgtcinsertpkg

\title{The Plausibility Paradox For Scaled-Down Users In Virtual Environments}

\author{Matti Pouke \thanks{e-mail:matti.pouke@oulu.fi}\\ %
        \scriptsize University of Oulu, Center for Ubiquitous Computing %
\and Katherine J. Mimnaugh\thanks{e-mail:katherine.mimnaugh@oulu.fi}\\ %
     \scriptsize University of Oulu, Center for Ubiquitous Computing %
\and Timo Ojala\thanks{e-mail:timo.ojala@oulu.fi}\\ %
      \scriptsize University of Oulu, Center for Ubiquitous Computing\and
Steven M. LaValle\thanks{e-mail:steven.lavalle@oulu.fi}\\ %
      \scriptsize University of Oulu, Center for Ubiquitous Computing}

\abstract{This paper identifies a new phenomenon: when users interact with simulated objects in a virtual environment where the user is much smaller than usual, there is a mismatch between the object physics that they expect and the object physics that would be correct at that scale. We report the findings of our study investigating the relationship between perceived realism and a physically accurate approximation of reality in a virtual reality experience in which the user has been scaled down by a factor of ten. We conducted a within-subjects experiment in which 44 subjects performed a simple interaction task with objects under two different physics simulation conditions. In one condition, the objects, when dropped and thrown, behaved accurately according to the physics that would be correct at that reduced scale in the real world, our \textit{true physics} condition. In the other condition, the \textit{movie physics} condition, the objects behaved in a similar manner as they would if no scaling of the user had occurred. We found that a significant majority of the users considered the latter condition to be the more realistic one. We argue that our findings have implications for many virtual reality and telepresence applications involving operation with simulated or physical objects in small scales.  %
} 

\CCScatlist{ 
  \CCScat{H.5.1}{Information Interfaces and Presentation}%
{Multimedia Information Systems}{Artificial, augmented, and virtual realities}
}

\begin{document}
\begin{NoHyper}


\maketitle

\section{Introduction} 
The question of how our body influences the way we perceive the world has been pondered in both century-old philosophical texts as well as in modern research. Indeed, many research studies have found a "body-scaling effect": if presented with mismatching size cues, humans tend to use their visible body as the dominant cue when perceiving sizes and distances \cite{banakou_illusory_2013, langbehn_scale_2016, linkenauger_welcome_2013, van_der_hoort_being_2011, ogawa2017distortion}. If one, for example, was somehow shrunk to the size of a doll, that person would be inclined to regard the world as scaled-up and him/herself as normal-sized \cite{van_der_hoort_being_2011}. Currently, not much is known about how such scaling down of oneself would affect a person's perception of physical phenomena, such as accelerations. Interestingly, if we pay attention to how scaled-down characters interact with their surroundings in many works of fiction, the tendency to represent the world as scaled up in comparison to normal-sized protagonists can be observed. Early examples can be seen in the classic film \textit{The Incredible Shrinking Man}. When the main character throws grains of sand off the table while insect-sized, the grains accelerate and fall as if they were boulders - when they should be falling down instantly. Similarly, when the character is awash by rainwater holding onto a pencil, the water and the pencil act more akin to a river and a log when the pencil should be bobbing with few waves and no visible whitewater should be apparent. While the deficiencies in the realism of the \textit{Incredible Shrinking Man} can be attributed to 1950s technologies, similar inaccuracies still remain in modern movies from \textit{Honey I Shrunk the Kids} to \textit{Downsizing}. These inaccuracies are not necessarily resulting from directors’ lack of understanding of physics, but might be conscious choices to represent what the viewers would expect. 

\begin{figure}[t]
\includegraphics[width=0.5\textwidth]{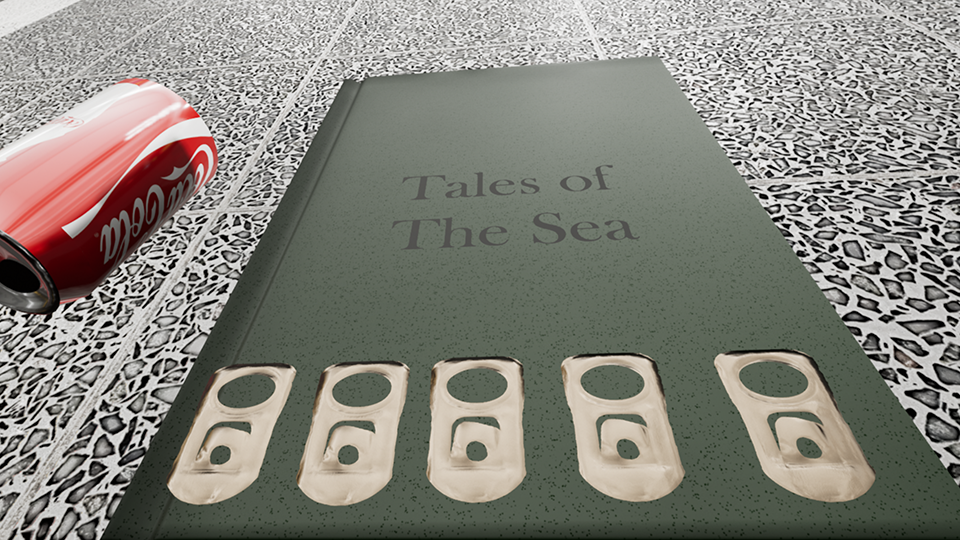}
\caption{First person perspective from the subject's point of view in the VE at the start of the experiment.}
\label{book}
\end{figure}

Virtual reality (VR) and telepresence applications allow humans to live through experiences such as the \textit{Incredible Shrinking Man} through the eyes of a scaled-down entity. A specific category of virtual environments (VEs) providing such experiences are multiscale collaborative virtual environments (mCVEs), in which multiple users can collaborate in, for example, architectural or medical visualizations across multiple, nested levels of scale (e.g., \cite{kopper_design_2006, zhang_mcves:_2005}). In addition, the scaling of users has been utilized in several collaborative mixed reality (MR) systems (e.g., \cite{billinghurst_magicbook:_2001, piumsomboon_superman_2018}). Teleoperation of robots can allow humans to interact with the physical world at micro- and nanoscale. Similar to mCVEs, robotic teleoperation systems that take place in multiple scales are beginning to emerge \cite{izumihara2019transfantome}. While teleoperation in the physical world can leverage stereoscopic camera systems resembling immersive VR applications \cite{hatamura_direct_1990}, purely virtual representations leveraging computer graphics can be used in, for example, educational and training systems for micro- and nanoscale tasks \cite{bolopion_review_2013, millet_improving_2008}. Robotic surgery systems can perform operations at a microscopic level \cite{hongo_neurobot:_2002} whereas stereoscopic VR can be utilized in telesurgery \cite{shenai_virtual_2014}. The benefits of VEs have been identified in various design and prototyping processes \cite{mujber_virtual_2004}; these processes can be extended into small-scale VEs as well. Already two decades ago, both the design \cite{li_virtual_2001} and assembly \cite{alex_virtual_1998} of microelectromechanical systems (MEMS) were prototyped through desktop VEs.

We believe that understanding human perception of scale-variable phenomena will be helpful for the future design of applications such as those listed above. While existing research has addressed many perceptual questions, such as the perception of distance and dimensions after altering one's virtual size (e.g., \cite{van_der_hoort_being_2011, banakou_illusory_2013, kim2017dwarf}), the perception of the behavior of physical objects has received relatively little attention. There are many potential future use cases for user scaling that might require interaction with physical or physically simulated objects. We, however, argue that it is not inherently intuitive for humans to perceive physical phenomena, such as rigid body dynamics, in scales that differ greatly from a normal human scale. An object dropped from 20cm takes significantly less time falling than an object dropped from 2m, and their perceived accelerations are different. Additional physical phenomena, such as fluid dynamics, frictions, and static electricity might affect interactions even further as the scale of the operations becomes smaller. For this reason, additional consideration is required when designing systems in which real or virtual interactions take place in atypical scales, and thus it is important to understand human perception regarding physical phenomena when scaled. In this paper, we present our early results investigating the aforementioned phenomenon. More specifically, we focus on the mismatch between perceived realism and a physically accurate approximation of reality when interacting in a VE while scaled down by a factor of ten. We hypothesize that people are not blind to changes in scale; however, when presented with two different scale-dependent rigid body dynamics approximations, they are more likely to consider the physically inaccurate one to be the more perceptually realistic one.

This paper progresses as follows. Section 2 presents previous research related to this work. Section 3 will outline our research method and describe the experimental setup. Section 4 will introduce our results. Section 5 discusses our findings while Section 6 concludes the paper.

\section{Related Research}
The manipulation of a user's scale can be accomplished by changing various properties of the virtual character the user is controlling in the VE. Changing these properties has various subjective effects. When scaling a user's virtual size, one of the most obvious properties to change is the viewpoint height, as it defines the virtual camera origin in relation to the VE, simulating a change in physical size. Viewpoint height affects egocentric distance perception \cite{leyrer_influence_2011, zhang_mcves:_2005}. Interestingly, minor changes in viewpoint height might go unnoticed by users \cite{leyrer_influence_2011,deng2019floating}. Users' interaction capabilities such as locomotion speed and interaction distance can be changed according to scale, depending on the purpose of the application \cite{zhang_mcves:_2005}. When using a head mounted display (HMD), the scaling of the user can also affect the virtual interpupillary distance (IPD), which is the distance between the two virtual cameras that are used to render the environment for the user. Changing this distance can affect the user’s sense of their own size relative to the VE \cite{piumsomboon_superman_2018, kim2017dwarf}.

Humans generally seem to utilize their own body as a primary metric for scale (an effect also referred to as \textit{body scaling}), and the virtual representation of the user's body greatly affects their perception of sizes and distances in the VE \cite{8798040, ogawa2017distortion}. Linkenauger et al. \cite{linkenauger_welcome_2013} studied the role of one's hand as a metric for size perception; they conducted an experiment where they scaled the users’ virtual hand and found out that it had a strong correlation with perceived object size. Ogawa et al. \cite{8798040} studied the effect of hand visual fidelity on object size perception and found that the visual realism of the hand affects the extent of the body scaling effect. van der Hoort et al. \cite{van_der_hoort_being_2011} embodied the entire user in a doll’s body as well as in a giant’s body using a stereoscopic video camera system and an HMD. They found that the embodiment significantly affected the users’ distance and size perceptions, especially if the user experienced a strong body ownership illusion \cite{slater2009inducing} with the virtual body. Banakou et al. \cite{banakou_illusory_2013} compared the effects of embodying the user as a child versus as a scaled-down adult. They found that the effect of altered size and distance perceptions was even larger when embodied as a child, and it also made the users associate themselves with childlike personality traits. 

The environment, whether real or virtual, affects the perception of scale. Humans generally underestimate egocentric distances in VEs, except when the VE is faithfully modeled to represent a real environment \cite{renner_perception_2013}. However, if a familiar room is scaled slightly up or down, underestimations are reintroduced \cite{thompson_elucidating_2007}. Familiar size cues also affect the sensitivity to eye height manipulations \cite{deng2019floating}. Langbehn et al. \cite{langbehn_scale_2016} studied the effect of body and environment representations as well as the scale of external avatars on users' perception of dominant scale in mCVEs (the dominant referring to the “true” scale in an mCVE system where users can coexist in multiple scales). They found that humans tended to use their body as the primary metric for judging their own size and the environment if the representation of one's own body was not available. In addition, an environment with familiar size cues helps in the determination of scale, while an abstract environment does not. They also found that the majority of subjects tended to estimate external avatars to be at the dominant scale instead of themselves.

Studies in micro- and nanoscale teleoperation have revealed that, due to changes in physics, interactions at these scales can become difficult for the human operator, but education inside virtual reality environments has been found to alleviate this drawback \cite{millet_improving_2008, sitti_microscale_2007}. Besides this work, there is little research on  human perception of physical phenomena when users are scaled-down in a normal-sized environment in VR.

\begin{figure}[t]
\includegraphics[width=0.5\textwidth]{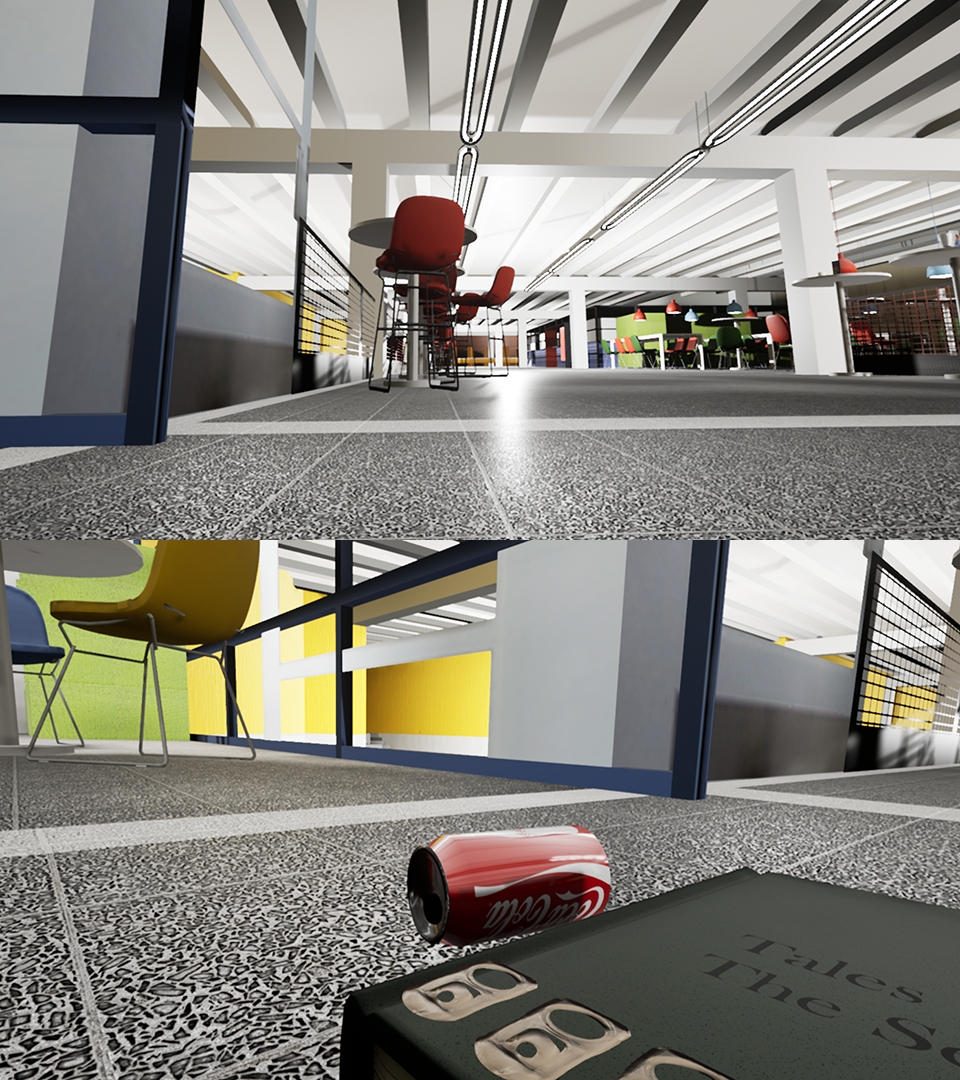}
\caption{A screenshot of the VE from the subject's perspective when looking forward and upward with book and tabs below the line of sight (Top) and when looking left (Bottom).}
\label{scene1}
\end{figure}

\subsection{Presence and Plausibility}
The concepts of immersion \cite{slater_framework_1997}, presence and plausibility \cite{slater_place_2009} are relevant for this study. In Slater’s classical definition, the level of immersion refers to the level of technical fidelity of the VR system (i.e., resolution, field of view, vividness of graphics) \cite{slater_framework_1997}. In addition, the realism of the user’s response to the VR system depends on two orthogonal components, presence or place illusion (PI) and the plausibility illusion (PSI) \cite{slater_place_2009}. PI refers to the sensation of being in another place, while PSI refers to the perceived believability of the virtual scenario or experience (illusion as being there vs. realness of what is happening) \cite{rovira_use_2009}. PSI depends on the extent to which the VE can produce authentic responses for user actions. Rovira et al. \cite{rovira_use_2009} argued that for PSI to occur, participants must perceive themselves as beings that exist in the VE; user actions must elicit actions in the VE and the VE must acknowledge the user (i.e., virtual characters react to the user). In addition, the VE should match the users’ prior knowledge and expectations \cite{rovira_use_2009}. Skarbez et al. \cite{skarbez_psychophysical_2017} used the term \textit{coherence} to refer to the aspects of a VE that contribute to PSI, such as virtual humans and the behavior of virtual objects. They argued that while immersion is a technical attribute that affects PI, coherence is a similar technical attribute affecting PSI. 

In this study, we used the concept of PSI to study human perception of the behavior of physical objects while the subject was scaled down and interacting in a normal-sized environment. However, we delimited virtual characters out from the scope of this study. Instead, we were interested in how subjects would perceive the coherence in terms of behavior of virtual objects, when it would be reasonable to expect a mismatch between expectations and correctly simulated reality. In addition, we investigate whether the extent of PI affects PSI in this particular context.

\section{Methods}
\begin{figure}[t]
\includegraphics[width=0.5\textwidth]{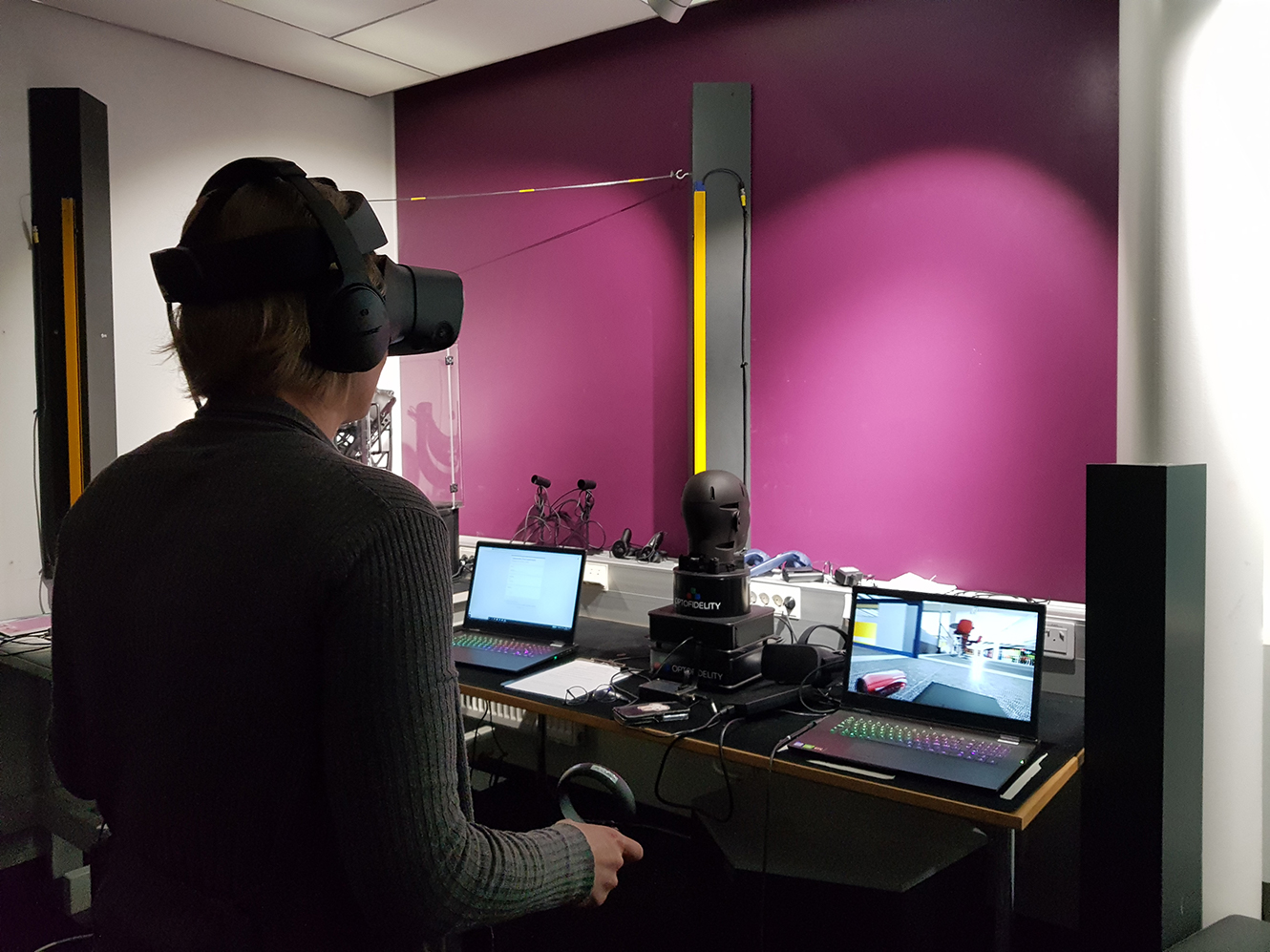}
\caption{Third person perspective of a subject using the experiment software while viewing the VE.}
\label{Experiment}
\end{figure}
The specific objective of this study was to investigate the PSI of subjects in two different physics conditions. The purpose of both conditions was to visually represent a scaled-down subject in a normal-sized environment, and the physics simulations differed between the conditions as follows. In the condition we call \textit{true physics}, the rigid body dynamics affect virtual objects in an approximately similar way to what would be accurate at that scale. In the \textit{movie physics} condition (named after physical behavior as typically seen in Hollywood movies in scenes depicting scaled-down characters), rigid body dynamics behave in what would be the approximation of a normal human scale. Our assumption was that the users would be able to distinguish between \textit{true physics} and \textit{movie physics}, and we predicted that subjects would be more  likely to feel and to expect the \textit{movie physics} condition to be the more perceptually realistic representation. This implies the Plausibility Paradox, a mismatch between perceived realism and the correct approximation of realism. 

\subsection{Hypotheses}
We hypothesized that in the \textit{true physics} condition, the behavior of physical objects would feel incorrect for subjects despite their knowledge of being virtually shrunk down. More specifically, our hypotheses were as follows:

\begin{itemize}
\item [H1:] For a scaled-down user, \textit{movie physics} is more likely to feel realistic than \textit{true physics}.

\item [H2:] For a scaled-down user, \textit{movie physics} is more likely to match a user's expectations than \textit{true physics}. 

\end{itemize}

\subsection{Experimental Apparatus}
We designed an experiment in which the subjects performed a simple interaction task in Unreal Engine 4.22 (UE) based VEs using the physics conditions described above. In both conditions, the scaling operations took place in one order of magnitude, giving the impression of a doll-sized perspective. We did not use full body tracking or attempt to induce a strong body ownership illusion \cite{slater2009inducing}, so there was no visualization of any body parts in the VE other than the subject's hands. We used the default UE VR hand visualization for interaction and to present a medium-fidelity body size cue \cite{8798040}. There was no difference between the conditions regarding how the hands functioned or how the user was able to move.  

To help in providing accurate size cues, we modeled the VE acting as the base for the experiment to resemble a location in the main corridor of the campus in which the study took place. The dimensions and materials of the VE were modeled using the real environment as the basis. In addition, we took measurements of various real-world objects, such as chairs, tables, and leaflets, which we modeled and scaled accordingly and placed in the VE as static objects.

The scaling of the user in the \textit{true physics} condition was achieved by shrinking the user with the UE built-in \textit{World to Meters} parameter, which automatically scales the player character's height, virtual IPD and interaction distance. The skeletal meshes representing the player character's virtual hands were scaled down manually. In the \textit{movie physics} condition, the player character properties were kept as default and the VE was scaled up instead. The purpose for this approach was to give the visual illusion of a scaled-down user, while retaining physics conditions that correspond to the normal human scale. The sizes and relative distances of scene objects were increased by a factor of ten. In addition, the properties of lights and reflection capture objects were adjusted so that the overall visual appearance of both conditions were kept as similar as possible.   

\subsubsection{Interaction Task}
The interaction task consisted of the manipulation of virtual soda can pull tabs approximately 3cm lengthwise and 1.9cm in width (as presented in Fig. \ref{book}). The tabs were chosen for the experiment both for their small, consistent mass as well as for being a reasonably authentic object that could be seen in the simulated VE. We considered a lightweight object to be most practical for simulating throwing in VR so that we would not have to simulate the decrease in hand acceleration due to increased inertia at the end of the arm or limitations due to arm strength \cite{cross2004physics}. In both conditions, the subjects would try dropping and throwing five tabs. Picking up and throwing the tabs took place utilizing the default mechanism in UE, similar to contemporary VR applications in general. The subjects simulated grabbing objects by squeezing the trigger of the motion controller and dropping them by releasing the trigger. Virtual throwing took place by swinging the motion controller and then releasing the trigger, and the object thrown retained its velocity at the moment of release, simulating throwing.

In the \textit{true physics} condition, the tabs would drop down fast, similarly as to if they were dropped from the height of 15-20 cm (simulated falling speed approximately 0.175s at 20cm in UE). In addition, the throwing distances would appear short because of the limited velocity that can be actuated due to real hand movements scaled down by an order of magnitude. The \textit{movie physics} condition, on the other hand, simulated the tabs as falling down more slowly, similarly to an object dropped from human height (simulated falling speed approximately 0.6375s at 2m in UE). In addition, the throwing distances were much larger in the \textit{movie physics} condition due to the larger velocity that the subjects were able to actuate on the tabs by virtual throwing. 

Due to the simulated size, the tabs were also different between conditions in terms of their bounciness (there were no changes in physics simulation properties, such as restitution). In the \textit{movie physics} condition, the tabs bounced visibly off surfaces, or jittered slightly after being dropped. However, in the \textit{true physics} condition, there was little to no visible bounciness.  

The tabs were placed on top of a large book so that the subjects would not have to pick them up from the floor. The book also provided an additional size cue. We gave the book a neutral, non-distracting appearance and a general title so that it was recognizable as a book, but would not otherwise draw too much attention. A Coca-Cola can was placed as a familiar sized cue on the left side of the book. Fig. \ref{book} shows the book and the tabs as seen in the beginning of the simulation. Fig. \ref{scene1} shows the scene as seen at the beginning of the simulation when looking forward (Top) and left (Bottom).

The virtual mass of the tabs was set at 1g in both conditions. Default physics settings in UE were utilized, with the exception of turning on the physics sub-stepping for additional physics accuracy by enabling physics engine updates between frames. Drag by air resistance was set to zero in both conditions. The simulation itself ran at stable 80 FPS which is the maximum frame rate of Oculus Rift S.

\subsection{Experimental Procedure}
The experiment was carried out as a within-subjects experiment, in which 44 subjects (23 females and 21 males) performed both conditions during one experiment. The order of the conditions was counterbalanced so that there was an equal number of male and female participants starting with each condition. The subjects' ages ranged from 19 to 66, mean and median ages being 30 and 26, respectively. The standard deviation for the ages was 10.4. The study was conducted either in English (12 females and 7 males) or in Finnish (11 females and 14 males), depending on the preference of the subject. 

The experiment was set in a laboratory in which the subjects used the Oculus Rift S system with provided Oculus Touch controllers for the experiment. The Rift S has a variable IPD software setting, so the IPD was set to 62.5 for females and 64.5 for males, the closest approximation possible based on the averages reported for adults \cite{dodgson2004variation}. In the beginning of each session, the subjects read through a written \textit{Information for Subjects} document and signed an informed consent sheet. The subjects were then instructed on using VR hardware, specifically how to use the Rift S Touch motion controllers for picking up and throwing objects. Next, they were instructed to stand on a particular starting spot in the laboratory (marked with masking tape), which was 110 cm away from the laptop used for the HMD. When the user was wearing the HMD and the motion controllers comfortably, the following instruction script was read in English or Finnish: "\textit{In this experiment, you are in a virtual reality environment, where you are at the university central hallway at night. You have been shrunk down to a size of a barbie doll, approximately 10-times smaller than your current height. You can move around a little bit by taking a few steps (but you don’t have to). You will see several pull tabs placed on a book in front of you. We would like you to pick one up and then let it fall to the floor. After that, we would like you to pick one up and throw it across the book in front of you. We would like you to try dropping and throwing the remaining pull tabs as well. After no pull tabs are remaining on the book, we will restart the experiment and ask you to repeat what you just did with the pull tabs. I will now put on the headphones, and then you may begin."}

Active noise-cancelling headphones were placed on the subject to block out any potential external noise from other rooms in the building, and then the experiment began. After performing both conditions, the headphones and the VR hardware was removed and the subject was asked to respond to a post-experiment questionnaire as well as a background questionnaire on a different laptop (seen left in Fig. \ref{Experiment}). The subjects were asked for any additional comments or questions, and if they could be contacted for future studies, and then given a gift certificate for two euros for participation. The average duration of the session was 20 minutes per subject.

\subsubsection{Questionnaires}
We collected plausibility related data using two forced choice questions (main questions 1 and 2), two open-ended questions (O1 and O2) and a 7-point  Likert scale questionnaire regarding the behavior of the tabs (L1-L5). In addition, the subjects filled out the extended version of the Slater-Usoh-Steed (SUS) Presence questionnaire \cite{slater_depth_1994, usoh2000using}, as well as a background information questionnaire. The main questions 1 and 2 were as follows:    

\begin{enumerate}
    \item \textit{Thinking back how the pull tabs were behaving in the experiment, which felt more realistic (like what would happen in the real world if you had been shrunk down), the first or the second time? }
    \item \textit{Thinking back how the pull tabs were behaving in the experiment, which matched your expectations (similar to what would happen in the real world if you had been shrunk down), the first or the second time?}

\end{enumerate}
The main questions were coupled with open-ended questions (O1 and O2), that were simply stated as \textit{"Why?"}. The purpose of the open-ended questions was to evaluate to what extent the subjects' responses were related to the physics or other reasons.

The forced-choice and open-ended questions were followed by a 7-point Likert scale questionnaire asking subjects to judge how they perceived various aspects related to the behavior of the tabs. Each question was stated twice in the questionnaire, referring to the first time and the second time subject interacted with the tabs (either using the \textit{true physics} and then the \textit{movie physics} or vice versa). The first three questions (L1-L3) were bipolar while the last two (L4, L5) were unipolar. The Likert questions L1-L5 and their associated scales were as follows:  

\begin{itemize}
    \item [L1] \textit{The falling speed of pull tabs (too slow, too fast)}
    \item [L2] \textit{The speed of pull tabs when thrown (too slow, too fast)}
    \item [L3] \textit{The distance of pull tabs when thrown (too close, too far)}
    \item [L4] \textit{The way the pull tabs were bouncing when thrown (incorrect, correct)}
    \item [L5] \textit{The impact of gravity on the pull tabs (incorrect, correct)}
\end{itemize}

All questions were presented in either English or Finnish, depending on which was chosen as the preferred language by the subject when signing up for the experiment.

\section{Results}
\begin{figure}[t]
\centering
\includegraphics[width=0.4\textwidth]{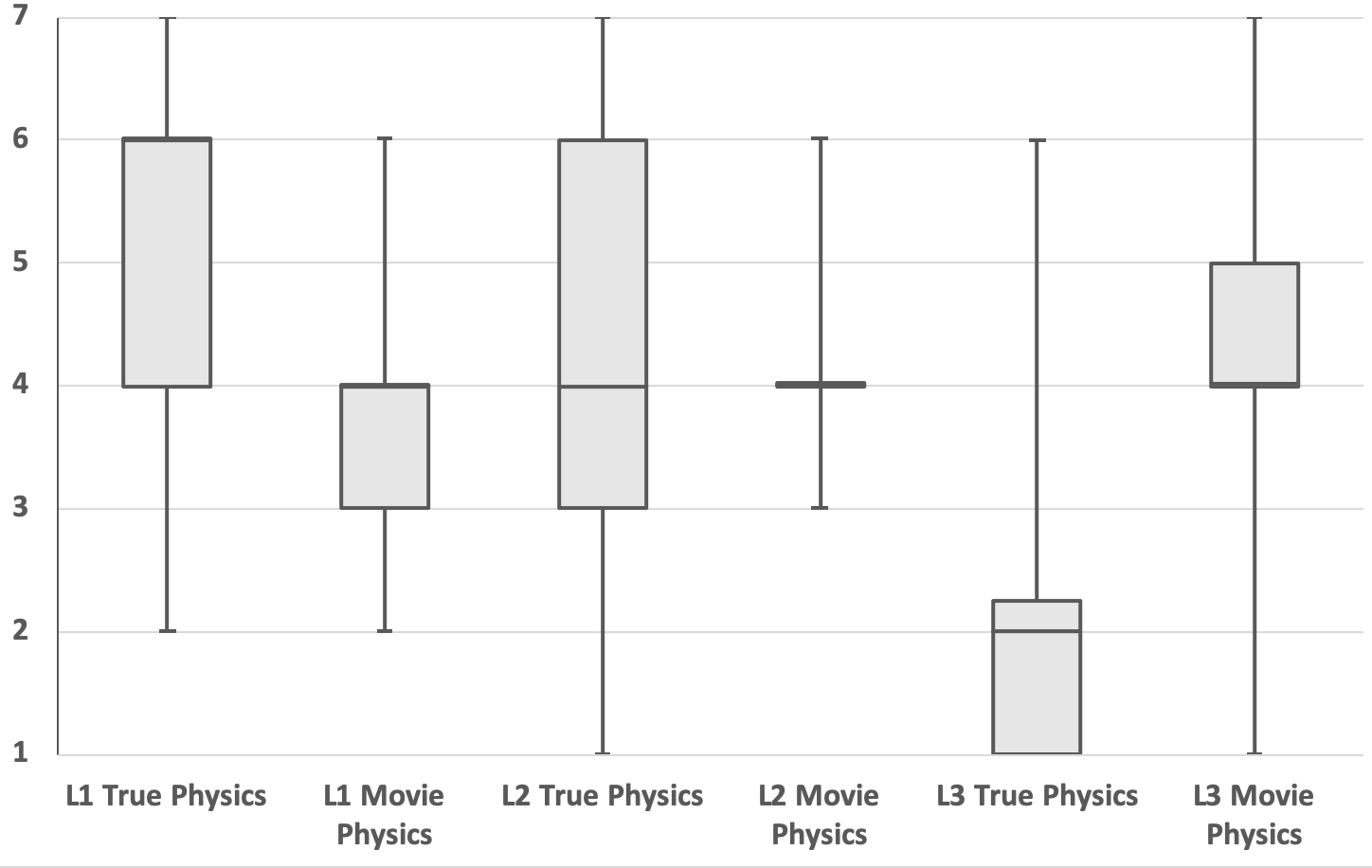}
\caption{Box plots visualizing the distribution of responses for questions L1-L3. Responses closer to 4 are perceived as closer to realism.}
\label{Likert13}
\end{figure}

Two subjects were removed from the analysis due to significantly different conditions as a result of issues with the functionality of the software or due to vision impairments. 

According to the responses to the main questions, the majority of the subjects considered the \textit{movie physics} condition as the more realistic one. Out of 44 subjects, 33 participants (73\%) responded to the first question that they considered the \textit{movie physics} condition more realistic, which confirms H1. For the second question, 42 out of 44 (93\%) subjects responded that the \textit{movie physics} matched their expectations better, which confirms H2. Furthermore, we analyzed the frequencies of responses to questions 1 and 2 with a binomial test and found their corresponding two-tailed p values as $p=0.004$  and $p = 1.7051^{-8}$ respectively. From this we can conclude that it is unlikely that the responses to questions 1 and 2 were due to chance. In addition, this implies that subjects were able to distinguish between the physics conditions and more consistently selected the \textit{movie physics} response, which was the inaccurate physics condition.

Out of twelve respondents who considered \textit{true physics} more realistic, nine responded that the \textit{movie physics} matched their expectations more. Only one subject considered the \textit{movie physics} more realistic while simultaneously stating the \textit{true physics} better matched their expectations. 

\subsection{Understanding the Contributing Factors}
We gathered supplementary data to further understand the results. These data include responses to open-ended questions O1 and O2, Likert-scale questions L1-L5, as well as subject background and self-reported level of presence.

\subsubsection{Open Ended Questions O1 and O2}
\begin{figure}
\centering
\includegraphics[width=0.4\textwidth]{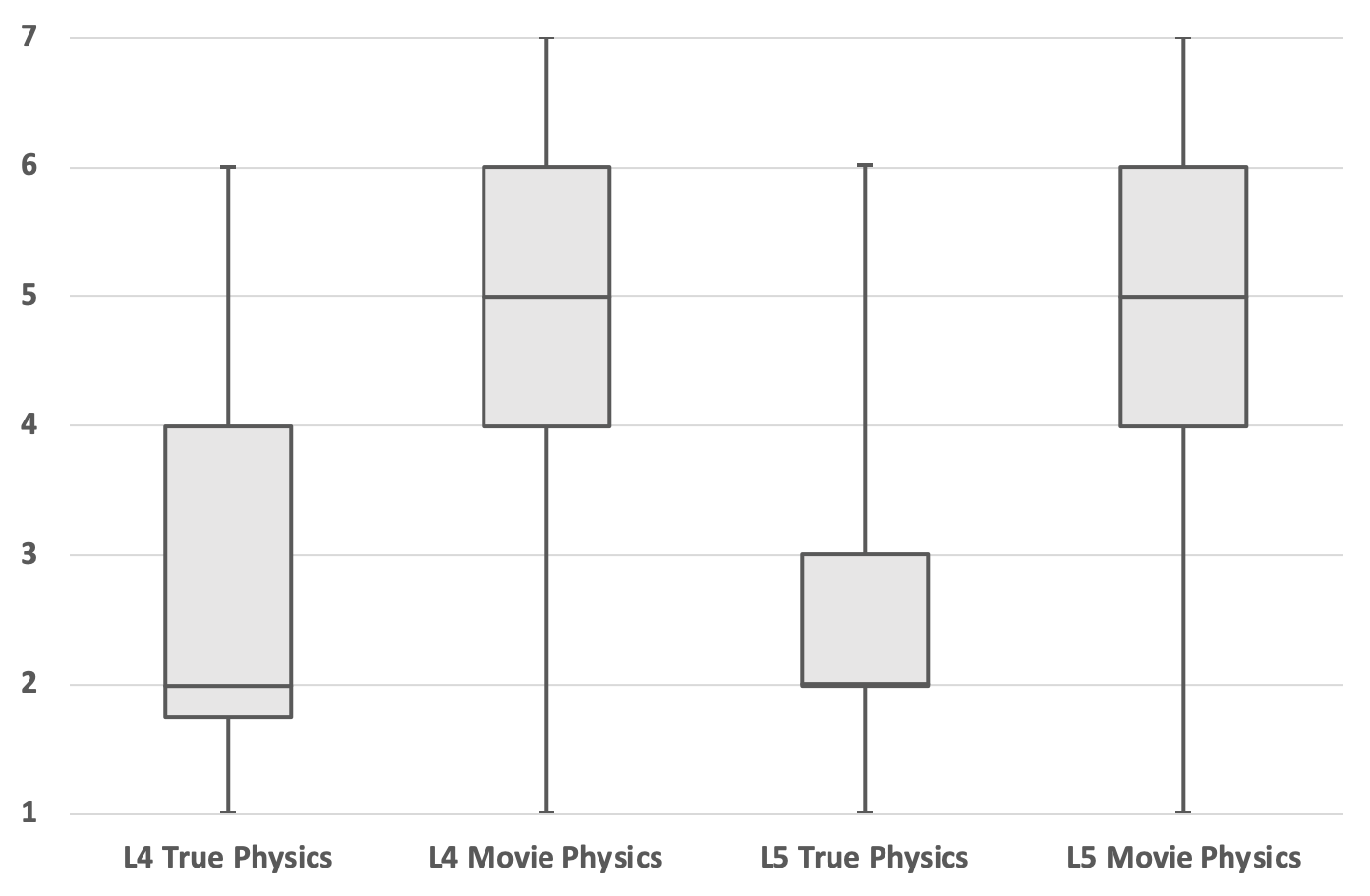}
\caption{Box plots visualizing the distribution of responses for questions L4-L5. Responses closer to 7 are perceived as closer to realism.}
\label{Likert45}
\end{figure}
The purpose of the open-ended questions was to evaluate to what extent the subjects' responses to the main questions 1 and 2 were related to the perceived realism of the physics. The responses consisted of one-sentence statements typed by the subjects. Thematic analysis with an inductive approach (e.g., \cite{patton2005qualitative}) was carried out independently by two researchers and used to identify categories in the response data. A summary of the responses is as follows:  

For O1, vast majority of the subjects (38 subjects) responded with a reason that can be attributed to the physics conditions (e.g., \textit{"Gravity feels more natural,"} \textit{"the tabs were flying in a more natural way"} or \textit{"Second time they felt too heavy"}). Five subjects also gave a secondary reason unrelated to physics (e.g., \textit{"movement in space felt more realistic, but the objects lacked 3D, ring pulls are not paper thin"} or \textit{"I am not sure but I think the second time they still moved a bit after I dropped them to the floor, before being completely still. I think I also managed to throw one of the pull tabs the second time, which felt more realistic than them dropping very quickly just right in front of me after I tried to throw them (but this could also just have been my inability to throw the first time)."}

Five subjects out of 44 gave a response related to general interaction or learning effect as the primary reason (e.g., "\textit{because I was more comfortable with the controllers after using them for some time, and I knew I could do more things now like throwing more far away after some time, and also they were moving more smoothly"} or \textit{"I'm kind of feel the same for both times. But maybe the second is more realistic just because I get use to it."}

Finally, one subject gave a reason completely unrelated to interaction (\textit{"the Coke can made the situation plausible}.") Observing recorded video material, it can be seen that this subject stopped to admire the Coca Cola can for a moment after one of the tabs landed close to it in the \textit{movie physics} condition. However, we do not know the exact reason why the Coca Cola can was chosen as the response. The cans were the same in both conditions so it may have been a more general comment. 

Open question O2 asked the subjects to report the reason for their choice of answer for the main question 2, asking which condition matched their expectations more. The subjects who had different responses for main questions 1 and 2 (5 subjects) reported their justifications for this (e.g., \textit{"I was not thinking I was shrunk. So it felt estrange to have such heavy pull tabs,"} \textit{"I didn't think at first (until I saw the previous question) shrinking down would also affect the time it takes for the objects to reach the ground. The physics first time behaved just like in normal life,"} or \textit{"Even though I knew I was shrunk down, I somehow could not think about it that way while doing the experiment"}).

The subjects who chose \textit{movie physics} (33 subjects) confirmed their answers for their first response by simply referring to their earlier response, or providing additional reasoning, such as \textit{"As I was taking a swing with my arms I was expecting them to land far away from me which they did only during the first time"} and \textit{"In the second time, the tabs were falling down surprisingly fast."}

The subjects who chose \textit{movie physics} while reporting reasoning unrelated to physics in O1 (4 subjects) mostly confirmed this reasoning in O2 as well. For example, \textit{"I was paying more attention to the behavior of the pull tabs, while in the first time the environment caught most of my attention"} (O1), \textit{"For the same reason as the previous response, although for some reason I liked the color of the pull tabs better in the first time, it was somehow more clear"} (O2). However, one subject referring to general interaction in O1 (\textit{"It felt easy to pick them up"}) gave another reason in O2 that seems more associated with physics, \textit{"First time throwing them felt more natural"} (O2).
One of the subjects preferring \textit{movie physics} expressed doubts for his responses to both questions with the statement in O2 as \textit{"The behavior seemed more natural, although probably the laws of physics tell otherwise."} The subject who responded regarding the soda can making the condition plausible in O1, gave a different type of reasoning in O2, \textit{"The tabs were flying plausibly. Especially one that even started gliding far away."}

In short, the responses to questions O1 and O2 indicate that majority of users (38 out of 44) made their choices primarily according to reasons related to the behavior of the physically simulated tabs. Other primary reasons were related to general interaction and learning effects. Four references were made to visual details as secondary reasons or general remarks (two references regarding appearance of tabs in O1 and two references to colors in O2).

\subsubsection{Likert Responses}
Inspecting the Likert responses for questions L1-L5, we found that the \textit{movie physics} condition was closer to perceived realism (median responses closer to 4 in questions 1 and 3 and closer to 7 in questions 4 and 5) in all questions except L2, in which the median response was the same for both conditions. We analyzed the responses to questions L1-L5 with the Wilcoxon Signed Rank test and found that the responses were significantly different (p \textless0.005) for all questions except L2, (p = 0.845). This gives additional confirmation that the subjects perceived the \textit{movie physics} condition more realistic due to differences in the behavior of the physically simulated tabs. A summary of responses including, median, mode and standard deviation for questions L1-L5 can be seen in Tab. \ref{Likertdata}. In addition, box plots visualizing the medians, interquartile ranges as well as minimum and maximum responses can be seen in Fig. \ref{Likert13} and Fig. \ref{Likert45}. 

\begin{table}[]
\small
\centering
\caption{Likert questionnaire data summarized. Responses perceived closer to realism are emphasized in bold.}
\begin{tabular}{|l|l|c|c|c|}
\hline
\textbf {Question} & \textbf {Condition} & \textbf{Median} & \textbf{Mode} & \textbf{STD} \\
\hline
\hline
\textbf{L1:} & \textit{True Physics} & 6 & 6 & 1.4 \\
Falling speed & \textit{Movie Physics} & \textbf{4} & \textbf{4} & 0.8         \\
\hline
\textbf{L2:} &  \textit{True Physics} & \textbf{4} & 6 & 1.8 \\
Speed when thrown & \textit{Movie Physics} & \textbf{4} & \textbf{4} & 0.9 \\
\hline
\textbf{L3:} &  \textit{True Physics}  & 2 & 2 & 1.3 \\
Distance when thrown & \textit{Movie Physics}  & \textbf{4} & \textbf{4} & 1.1 \\
\hline
\textbf{L4:} & \textit{True Physics}  & 2 & 2 & 1.6 \\
Bounciness & \textit{Movie Physics}  & \textbf{5} & \textbf{6} & 1.5 \\
\hline
\textbf{L5:} & \textit{True Physics} & 2 & 2 & 1.4 \\
Gravity & \textit{Movie Physics} & \textbf{5} & \textbf{6} & 1.4 \\
\hline
\end{tabular}
\label{Likertdata}
\end{table}


\subsubsection{Self-reported Presence}
Thirty-six out of 44 subjects had an SUS count larger than 0 (the number of responses per subject with a score of 6 or 7  \cite{slater_depth_1994}), with the median SUS count being 3. Similar to the study by Skarbez et al. \cite{skarbez_psychophysical_2017}, we divided the subjects into groups of low presence (SUS count 0-2) and high presence (SUS count 3-5). This split divided the population almost evenly with 23 subjects experiencing high presence and 21 experiencing low presence. This implies that 82\% of subjects experienced PI to at least some extent while 53\% experienced a high sense of presence.

We examined whether high or low presence and perceived realism are independent. Examining the presence groups, we found that their responses to main question 1 were almost equally distributed: 7 subjects out of 23 (30\%) from the high presence group chose \textit{true physics}, while this was true for 5 out of 21 (24\%) in the low presence group. We used the Fisher's Exact Test for independence to confirm that belonging to a group of high or low presence and the response to main question 1 are independent (p = 0.7). Thus, high or low presence did not affect the perception of realism.

\subsubsection{Effect of Background and SUS scores}
Furthermore, we used a binary logistic regression to analyze the effects of subject background and presence on their responses to main question 1. We used \textit{Educational Background, Gender, Age, VR Experience, Gaming Experience, SUS Average and SUS Score} as independent variables and the response to main question 1 as the dependent variable. 

For analysis purposes, we transformed the Background Questionnaire responses to \textit{Educational Background } into a binary variable consisting of roughly equal sized groups of \textit{Natural Sciences and Engineering} (25 subjects) and \textit{Social Sciences} (19 subjects). In addition, the open responses to \textit{VR Experience} and \textit{Gaming Experience} was transformed into respective ordinal variables ranging from 0 (no experience) to 4 (plenty of experience). When interpreting the \textit{Gaming Experience} responses, additional emphasis was given to recent experience as well as experience regarding PC and console based 3D gaming (such as first person shooters and simulators) due to the tendency of such games to contain game physics simulations similar to those used in this experiment. The responses to SUS scores were transformed into two ordinal variables consisting of average of responses as well as the computed SUS count. 

The logistic regression model was unable to predict the response using the independent variables. The model explained 17\% of the variance (Nagelkerke's $R^2$) in perceived realism. Although the overall classification rate was 72.7\%, only 16.7\% (two responses) of the \textit{true physics} responses were correctly classified. None of the independent variables had a significant effect on the prediction of the response (p = 0.184 - 0.858). According to this analysis, the perception of realism was not significantly affected by background, education or gaming in our subjects. The level of presence according to self-reported SUS score did not have any effect either.

\subsubsection{Perception of Mass and Strength}
Although we never queried subjects directly regarding the physical properties of the tabs themselves, several subjects commented on the weight of the tabs or their own strength when interacting with the tabs. Five of the subjects who responded in English commented on the feeling of the perceived heaviness of the tabs. For example, in responding to why they selected the \textit{movie physics} as more realistic, one subject commented, \textit{"The pull tabs looked and felt heavier and were easier to throw, as I would expect."} A second subject also commented that the pull tabs in the \textit{movie physics} condition felt heavier, \textit{"They fell in the right place, they had weight and they flew in a realistic projectory [sic]."} However, another subject, remarking on why the \textit{movie physics} was more realistic, said,  \textit{"Second time they felt too heavy,"} referring to the perceived increase in weight of the tabs during the \textit{real physics} condition. It is interesting to consider these spontaneous responses regarding differences in the weight of the tabs given than there was no change in the controllers that the subjects used for each condition. Subjects held the controllers through the whole experiment and while the condition was changed without ever setting them down. This could be an indication of a pseudohaptic effect \cite{lecuyer2009simulating} (for example, manipulating the control-to-display ratio of the visual feedback when lifting an object can give the user an illusion of increased weight \cite{samad2019pseudo}). However, it is possible that the subjects were simply referring to the visible trajectories and falling speed of objects (as in the tabs \textit{seemed} heavier instead of tabs \textit{felt heavier}). Several of the responses in Finnish specifically contemplated the assumed weight of the tabs in regards to how more much power they would have needed to use to throw the tabs given their reduction in size. For example, \textit{"Tabs should probably fly a little further and not just fall down after throwing even though I was doll sized, because tabs are aluminium and weigh practically nothing"}.

\section{Discussion}
The results imply that we have identified a strong paradox concerning PSI in VEs in which the user has been scaled down. According to the results, almost a 3/4 majority (73\%) of the subjects found the \textit{movie physics} condition to be more realistic. In addition, a 9/10 majority (91\%) of subjects considered the \textit{movie physics} condition as better matching their expectations. From this, we conclude that even subjects who believed the \textit{true physics} to be a correct representation of reality still considered it to be surprising. This reasoning was also often present in the responses to open questions O1-O2. According to O1-O2, almost all of the subjects considered the perception of realism to be related to the physics behavior of the tabs. In addition, a small number of subjects gave responses related to general interaction reasons, including learning how to use the controllers correctly. A few secondary reasons or remarks were made referring to a scene object or other visual details. According to the responses to O2, most of the subjects who preferred \textit{true physics} as the realistic one stated that during the experiment it was still difficult to understand why the physics functioned the way that it did. 

We used the Likert scale questionnaires to gather additional insights and confirmation for our results. The questions were focused on various dynamic properties of the tabs so that we could more specifically pinpoint the effects of physics simulations on perceived realism. These responses imply preferences towards \textit{movie physics} as well, with significant differences regarding the perceived realism of the tab behavior (with the exception of question L2). However, in this question as well, the most popular responses indicated a preference for the realism of the \textit{movie physics} (see Tab. \ref{Likertdata} and Fig. \ref{Likert13}). The Likert responses give us additional confirmation that physically accurate representations of the physics during scaled-down interaction in VEs are not inherently intuitive for users. According to the results, accurate accelerations and falling speeds of objects were perceived as unrealistic. The distance that the subjects were able to throw the tabs was seen mostly as too short (although there were also responses that considered the \textit{movie physics} enabling too far throwing distances even if \textit{true physics} was considered short). In addition, responses regarding the bounciness of the of the tabs imply that subjects expected the tabs to behave similarly as if they were enlarged 10-fold.

We inspected the effects of various aspects of the subjects' background on their responses to O1. It could be that that subjects with a knowledge of physics, for example, might prefer the \textit{true physics} condition. However, we found no such effects in our subject group. In addition, we did not find the self-reported level of presence \cite{slater_depth_1994}, either as SUS counts or by dividing subjects into groups of high and low presence, to affect the response to O1 in our subject group.

\subsection{Implications}
In this paper, we introduce the Plausibility Paradox concerning small-scale interactions in VEs - when the expectations of a user do not match with reality. We argue that this finding has potential future implications to VR and telepresence applications. Through recent advances in consumer VR hardware as well as sub-microscopic \cite{plisson_2d_2015} and even atomic \cite{zheng_motioncor2:_2017} level imaging techniques, it is possible that we will witness an increasing exploitation of scaled-down VR interaction in the future. These utilizations could potentially include commercial systems outside of the scientific domain, such as with teleoperated maintenance robots or commercial virtual design solutions at a microscopic scale. However, at this stage, it is not known whether it is intuitive for humans to operate at small scales, especially if it involves operating in the real world or with realistically simulated physics. As can be seen by our initial results, the perception of physical phenomena as a scaled-down entity is likely to be unintuitive for most (it was interesting to note, however, that half of the subjects experienced a strong PI despite the apparent improbability of the experience of being doll-sized). As the scale of operation decreases, perceived frictions and accelerations increase, which has already been found problematic for humans in robotic micro- and nano-level operations \cite{sitti_microscale_2007}. As the scale decreases further, these perceived distortions amplify, and additional phenomena, such as fluid dynamics and static electricity, come into play as well. Relative changes in the environment would also provide additional challenges in the physical domain. For example, a floor that is  effectively smooth on a regular scale might become bumpy and full  of cracks. Grit and dirt might become actual obstacles for  navigation. Vibrations from passersby that would be otherwise indistinguishable might feel like earthquakes. 

We argue that these challenges provide interesting avenues for future VR research. VR education has already been seen as a potential remedy for  some issues of small-scale activities in the field of teleoperation \cite{millet_improving_2008}. 

\subsection{Challenges and Limitations}
An obvious outlier in the responses was the question L2 (see Fig. \ref{Likert13} and Tab. \ref{Likertdata}). Whereas in the other questions, the responses seem relatively consistent indicating a stronger preference towards one condition or the other, L2 is an exception. Inspecting the distribution of responses in question L2, it can be seen that the \textit{true physics} condition contains responses that are rather uniformly distributed in comparison to the \textit{movie physics} condition; the STD in the \textit{true physics} condition is twice as large as in the \textit{movie physics}. Whereas responses the L2 \textit{movie physics} condition was considered realistic (4, neither too fast nor two slow) by the vast majority, the \textit{real physics} condition received an almost equal number of responses between 2 (too slow) and 6 (too fast). We suspect that the uncharacteristic distribution of responses might be due to a poor wording in the question L2, \textit{The speed of pull tabs when thrown}. While we tried to ask how the subjects perceived the time of flight of the tabs, it could be that subjects had other interpretations for the question resulting in inconsistent responses. Similar inconsistency was found in responses from both Finnish and English speaking subjects, so we do not think the confusion was specific to the words themselves for either version. Rather, we speculate that some subjects thought we meant the speed of the tab in leaving their hand (resulting in short flight distance) when they threw, and others thought we meant the speed that the tab moved through the air. Alternate interpretations could have resulted from misinterpreting the action of the tabs as having been caused by their own inability to throw the tabs correctly.

According to both verbal comments during the experiment as well as responses to questions O1 and O2, some of the subjects starting with the \textit{true physics} condition thought that the reason for their difficulty in throwing the tabs to a far distance was a result of their own inability to use the controllers and not related to aspects of the environment. Although some subjects realized during the subsequent \textit{movie physics} condition that the behavior of the tabs was an experimental manipulation and not due to their own failure, there were still three subjects that stated as their main reason for preferring the \textit{movie physics} condition to be the fact that they had learned how to use the controllers. For subjects that had the \textit{movie physics} first, there did not seem to be any ambiguity that the difference in the behavior of the tabs was related to the environment. While a training session helping to learn the controllers might have been helpful, we believe that it could have introduced unwanted priming for subjects regarding the expected behavior of physics.

Another obvious limitation is the fact that it is currently difficult to realistically simulate object mass in VR. While we chose the soda can pull tabs for the task partly because of their light mass, there was some speculation among responses to O1-O2 on whether the weight of the object and/or simulated arm strength affected object manipulation.

During a few of experiment sessions, there were occurrences which could have broken presence or caused differences in the experiences of the participants. Two subjects became very active in the VE and accidentally bumped into furniture in the experimental room. For two subjects, a physics engine bug caused a single tab to land in an unrealistic orientation during the \textit{true physics} condition. For one subject trying to throw the tab with two hands, a bug caused the tab to catapult unrealistically far. We are not sure to what extent the subjects noticed these bugs or if it affected their responses. Additionally, although we tried to keep the visual appearances of the two conditions as similar as possible, the differences in the environment scale in the UE to simulate the two types of physics led to very subtle differences in brightness between the two conditions. Though we were initially of the impression that the differences were nearly impossible to distinguish, there were two responses to O2 that commented on differences between the visual appearance of the conditions.

Finally, there were subjects who were not always paying close attention to the flying or falling characteristics of the tabs, or did not wait until the reading of the instructions was finished. Not observing the tabs properly might have introduced inaccuracies to their responses. This came up with both verbal comments after the experiment as well as responses to O1-O2.

\section{Conclusion and Future Work}
In this paper, we present a novel phenomenon regarding the plausibility of physical interactions for scaled-down users in normal-sized VEs; when users interact with physically simulated objects in a VE where the user is much smaller than a regular human scale, there is a mismatch between the object physics that they expect and the object physics that is the correct approximation of reality at that scale. We argue that this finding opens many interesting avenues for future research regarding mCVEs, scaled-down user VR applications in general, as well as telepresence and teleoperation taking place on a reduced scale. Although the Plausibility Paradox discussed here is specifically related to small-scale users in normal-sized environments, there are most likely other situations in VR which similar mismatches can exist. 

In the future, we intend to focus more on the body scaling effect and its influence on interactions with physically simulated objects. We consider scales smaller than 1 order of magnitude interesting since we expect them to provide even greater plausibility mismatches in physical interactions. We will also seek to confirm the existence of our finding outside VR, for example using robotic teleoperation or telepresence in small scale. Finally, the subjective perception of weight that appeared in some of the open-ended responses can provide an interesting avenue for future research.  

\acknowledgments{
The authors wish to thank all the subjects for their participation in this study. This work was supported by the COMBAT project (293389) funded by the Strategic Research Council at the Academy of Finland and the PERCEPT project (322637) funded by the Academy of Finland, as well as the HUMORcc (6926/31/2018) funded by Business Finland.}

\bibliographystyle{abbrv-doi}

\bibliography{Aof2019}
\end{NoHyper}
\end{document}